\documentclass[fleqn,usenatbib]{mnras}

\usepackage{newtxtext,newtxmath}
\usepackage{gensymb}
\usepackage[T1]{fontenc}

\DeclareRobustCommand{\VAN}[3]{#2}
\let\VANthebibliography\thebibliography
\def\thebibliography{\DeclareRobustCommand{\VAN}[3]{##3}\VANthebibliography}

\usepackage{graphicx}	
\usepackage{subcaption} 
\usepackage{amsmath}	
\usepackage{xcolor} 

\title[Investigating the effects of fresh gas on the AGN luminosity]{Investigating the effects of fresh gas on the Active Galactic Nuclei luminosity of early- and late-type galaxies}

\author[Martyna W. Winiarska et al.]{
Martyna W. Winiarska$^{1,5}$\thanks{E-mail: martyna.w.winiarska@durham.ac.uk}, Sandra I. Raimundo$^{1,2}$, Timothy A. Davis$^{3}$, Rog\'erio Riffel$^{4}$, Francesco Shankar$^{1}$,\newauthor Phil Wiseman$^{1}$
\\
$^{1}$Physics and Astronomy, University of Southampton, Highfield, Southampton, SO17 1BJ, UK\\
$^{2}$DARK, Niels Bohr Institute, University of Copenhagen, Jagtvej 155, Copenhagen N, 2200, Denmark\\
$^{3}$Cardiff Hub for Astrophysics Research \& Technology, School of Physics \& Astronomy, Cardiff University, Queens Buildings, The Parade, Cardiff, CF24 3AA, UK\\
$^{4}$Departamento de Astronomia, Instituto de F\'\i sica, Universidade Federal do Rio Grande do Sul, CP 15051, 91501-970, Porto Alegre, RS, Brazil \\
$^{5}$Institute for Computational Cosmology, Department of Physics, Durham University, South Road, Durham DH1 3LE, UK \\
}

\date{Accepted XXX. Received YYY; in original form ZZZ}

\pubyear{2015}

\begin{document}
\label{firstpage}
\pagerange{\pageref{firstpage}--\pageref{lastpage}}
\maketitle

\begin{abstract}
The main fuelling processes for Active Galactic Nuclei (AGN) are currently unknown. Previous work showed that galaxies with a large kinematic misalignment between their stellar and gas reservoirs have a higher AGN fraction than galaxies without misalignment. Such misalignment is a strong indication of a past galaxy interaction or an external accretion event. In this work we use integral field spectroscopy data from the SAMI and MaNGA surveys to investigate the AGN luminosity as a function of kinematic misalignment angle. Our sample of AGN exhibit bolometric luminosities in the range $10^{40}$ to $10^{43}$ erg\,s$^{-1}$, indicative of low to moderate luminosity AGN. We find no correlation between AGN luminosity as a function of misalignment for AGN host galaxies from both surveys. We find some differences between the AGN luminosity of early- and late-type AGN host galaxies (ETGs, LTGs).  AGN in LTG hosts have a wider luminosity range, with most LTG hosts showing aligned stellar to gas kinematics. AGN in ETG hosts have a luminosity range that does not depend on misalignment angle, suggesting AGN in ETG hosts are consistent with being fuelled by external accretion events, irrespective of their stellar to gas kinematic misalignment. While all the AGN in ETGs in our sample are consistent with being activated and fuelled by external gas, the range of observed AGN luminosities is likely caused by secondary factors such as the amount of fresh gas brought into the galaxy by the external interaction.  \end{abstract}




\section{Introduction} \label{Intro}
It is now widely accepted that most galaxies host a supermassive black hole (SMBH) at their centres (e.g., \citealt{kormendy_richstone}, \citealt{macchetto}, \citealt{kormendy}). When these SMBHs are actively accreting gas and and emitting energy over periods of about 10$^{5}$ to 10$^{8}$ years (e.g., \citealt{woltjer}, \citealt{hoyle}, \citealt{lynden-bell}), they are referred to as Active Galactic Nuclei (AGN). To sustain their active phases, AGN require a continuous supply of fuel throughout their lifetimes (e.g., \citealt{combes}, \citealt{2019NatAs...3...48S}, \citealt{2024Galax..12...17H}), although it is still unclear how this fuel supply gets to the SMBH. The two main ways for this fuel supply are thought to come from mergers and accretion; the fuel could therefore have several origins, from recycled gas produced by stellar evolution within the host galaxy, to leftover gas from old stellar populations or supernova explosions, to external gas acquired through mergers (e.g., \citealt{2019NatAs...3...48S}, \citealt{2024Galax..12...17H}, \citealt{choi}, \citealt{Riffel+24}, \citealt{Rembold+24}). 

The fuelling mechanisms responsible for AGN fuelling are still poorly understood, with several possible origins for the gas, which we will briefly describe below. AGN fuel determines how strong the AGN can be, and therefore impacts AGN feedback which, in turn, affects the host galaxy. Theoretical models seem to favour secular processes as a way of fuelling the SMBH \citep{2019NatAs...3...48S}. This recycled gas could trigger both AGN and star formation, and could be brought into the galactic centre by winds moving it away from young stars or supernovae. In simulations from \citealt{choi}, recycled gas accounts for the largest gas source for SMBHs \citep[see][for an observational example]{Riffel+24}. However, this result is highly dependent on the galaxy's history - major and minor mergers also provide a substantial gas contribution to a galaxy (e.g., \citealt{Hopkins2008}, \citealt{choi}). They can provide both the SMBH growth fuel (external gas) and a way for the gas to channel towards the galaxy's centre due to, for example, cloud-cloud collisions \citep{shlosman}. External gas accounts for either the first or second most important gas contribution to SMBH growth in simulations \citep{choi}. 

The AGN fuelling process is difficult to study due to the scales involved \citep{2019NatAs...3...48S}; the AGN host galaxy can be many tens of kiloparsecs across, while the accretion disc is the size of a few light days \citep{peterson}. Therefore, it is difficult to resolve how the gas could flow into the SMBH to sustain its activity \citep{shlosman}. Very high angular momentum loss is needed for the gas to flow to the galaxy centre; a possible way to do this could be an external process like a merger (e.g., \citealt{shlosman}, \citealt{lagos}). Collisions between existing gas in a galaxy (native gas) and newly acquired gas (fresh gas) through mergers can potentially lead to angular momentum cancellation (e.g. \citealt{shlosman}, \citealt{capelo}, \citealt{jin}), which could promote the gas inflow to the galaxy centre to fuel the AGN. There is in fact observational evidence supporting this, at least for the most luminous AGN which seem to favour galaxy merger systems (e.g., \citealt{urrutia}, \citealt{fan}). In practice, SMBHs most likely receive their gas from multiple sources  - external accretion events, recycled gas, stochastic gas infall and leftover gas (e.g., \citealt{navarro}, \citealt{2024Galax..12...17H}, \citealt{choi}). This is supported by the fact that in simulations not all merger events result in an AGN (\citealt{Kocevski2015}, \citealt{choi}). 

The presence of kinematic misalignment in a galaxy has been suggested as evidence for accretion of external gas into said galaxy (e.g. \citealt{bertola}, \citealt{davis_bureau}). Misalignment in a galaxy is defined as the offset between the main kinematic axes of rotation of stars and gas. Simulations that investigate misaligned galaxies find that counter-rotating (misalignment angle $ = 180^\circ$) or misaligned galaxy structures (for example a misaligned gas disc) can promote gas inflow into the SMBH (e.g. \citealt{thakar}, \citealt{taylor}, \citealt{negri}, \citealt{capelo}). The externally accreted gas can lose angular momentum due to dynamical interactions with the native gas or stars, which could cause the accreted gas to flow towards the centre of the galaxy. In the simulations of \cite{taylor}, the time it takes for this to happen ranges from  $250$ Myr to $3$ Gyr, which could vary due to different gas abundances in different galaxy types. In the simulations of \citealt{choi}, there is a significant time delay between the external gas accretion and the start of AGN activity, averaging to about $1.85$ Gyr. These results could indicate that the timescale in which the acquired gas is available is long enough to sustain an AGN for its lifetime and to produce misaligned structures for a sufficiently long enough time for us to observe them. 

Therefore, studying kinematic misalignment could allow us to investigate whether the newly acquired gas of a galaxy is related to how its AGN is fuelled. \cite{sandra} find there is a higher fraction of AGN in galaxies with misalignment, suggesting that misalignment could lead or be related to the AGN fuelling process. There appears to be a clear divide between early-type galaxies (ETGs) and late-type galaxies (LTGs) when it comes to misalignment; the gas and stars in LTGs are mainly kinematically aligned, while gas and stars in ETGs with gas are often kinematically misaligned (e.g. \citealt{davis_alatalo}, \citealt{bryant_croom}, \citealt{sandra}). This difference could arise from the differing native gas abundances between ETGs and LTGs - early-types contain significantly less gas than late-types (at present time; e.g., \citealt{gallagher}, \citealt{lianou}, \citealt{Gobat_2018})., making it easier to form misalignment due to lower amount of angular momentum needed to change the spin axis. We shall therefore take morphology into consideration in this study of AGN fuelling (see also \citealt{shlosman, zhi_li}). 

In this paper, we aim to address if galaxies with kinematic misalignment have substantially different AGN luminosities. The results could allow us to narrow down the possible origin of AGN fuel. We investigate $\sim 230$ galaxies with spatially resolved maps of ionised gas and evidence of AGN activity and determine their distribution of AGN luminosity as a function of kinematic misalignment angle. In section \ref{data} we discuss the integral field spectroscopy data used. In Section \ref{methods}, we describe the methods for obtaining the AGN luminosities used for analysis, and present our relations between AGN luminosity and kinematic misalignment in Section \ref{results}. In Section \ref{discussion} we discuss our results in light of a possible AGN fuelling process. Throughout this paper we used concordance cosmology with the Hubble constant value of $73$\,km\,s$^{-1}$\,Mpc$^{-1}$ \citep{riess}.

\section{Data Analysis} \label{data}
The sample of AGN we use in this work comes from two integral field spectroscopy surveys: the Sydney-Australian-Astronomical-Observatory Multi-object Integral-Field Spectrograph (SAMI) Galaxy Survey Data Release 3 and the Mapping Nearby Galaxies at APO (MaNGA) Survey. Below we describe both surveys in more detail.

\subsection{SAMI sample} \label{imaj_quality_cut}
The SAMI Galaxy Survey observed $\sim$3000 nearby galaxies (target redshifts $0.004<z<0.095$) using integral field spectroscopy. The SAMI instrument covers the wavelength ranges of $3750-5750\textup{~\AA}$ and $6300-7400\textup{~\AA}$ with a blue and a red arm. The resolutions of the blue and red arm are $R\sim1700$, and $R\sim4500$, respectively \citep{SAMI_swoi}. The kinematic misalignment angles are calculated by \cite{sandra} (see Section \ref{misalignment_angle} for more details); redshifts and stellar masses for the sample used in this paper come from SAMI DR3 and are presented by \cite{croom}.  SAMI provides flux maps and noise maps \citep{SAMI_swoi} for a few emission lines, which have spaxels (spatial pixels) of size $0.5$ x $0.5$ arcsec. To trace the ionised gas distribution and properties we used the emission line flux maps for the galaxies from SAMI. The SAMI survey provides different flux maps created with different assumptions on the number of components used to fit each emission line. We used the \texttt{`default-recom'} data; \texttt{`default'} meaning the data cubed is not binned which ensures all available pixels are used, which maximises the spatial resolution, and \texttt{`recom'} indicates the fit with the recommended number of line components (which could be larger than one). This is a more accurate representation of the flux, as it is not relying on an assumed a peak shape. 

We used maps of emission line fluxes from H$\alpha$ $\lambda6563\textup{~\AA}$, H$\beta$ $\lambda4861\textup{~\AA}$ and $[\ion{O} {III}]$ $\lambda5007\textup{~\AA}$ transitions for 87 SAMI galaxies with AGN (confirmed by \citealt{sandra} through optical emission line ratios or AGN identification in the X-ray, radio or infrared).  $[\ion{O} {III}]$ emission lines are commonly used to detect AGN and can be used as a proxy for AGN luminosity (Section \ref{AGN_lums}). This transition has a high ionisation potential which requires energetic photons to be excited, that mostly come from the accretion disc of an AGN. The H$\alpha$ and H$\beta$ emission lines were used for computing the Balmer decrement which allowed us to estimate dust corrections for the  $[\ion{O} {III}]$ luminosities to obtain the bolometric luminosity range of our AGN. These are an approximation, since there may be off-nucleus, non-AGN minor contributions to the H$\alpha$, H$\beta$ and  $[\ion{O} {III}]$ emission lines fluxes.

We introduced our own quality cut on the SAMI flux maps. Each pixel used to calculate the integrated flux must have $S/N \geq 2$ on all flux maps (meaning noise can be $50\%$ of the signal value at most). This is in addition to SAMI quality cuts (each H$\alpha$ flux component must have $S/N \geq 5$ \citep{SAMI_swoi}, and each pixel on a flux map must have $S/N \geq 3$ per $\textup{~\AA}$ \citep{croom}). Such a stringent quality cut was to ensure both the reproducibility of our results and the presence of sufficient AGN emission, such that at least 4 pixels displayed it. This reduced our SAMI galaxy sample size to 86 galaxies. 

The SAMI survey provides visual morphology classifications based on indications of star formation, the presence of spiral arms and a bulge component \citep{SAMI_swoi}. There are four main galaxy types considered: ellipticals, S0s, early- and late-spirals, with intermediate stages where galaxies were difficult to classify and hold qualities of both groups, for example elliptical/S0s. 

\subsection{MaNGA sample} \label{manga_sample}
The MaNGA Galaxy Survey investigated a much larger galaxy sample of $\sim$10 000 nearby galaxies (target redshift $0.01<z<0.15$),  using IFU 
spectroscopy with wavelength range $3600-10400\textup{~\AA}$, with resolution $R\sim2000$ \citep{Abdurrouf+22,Aguado+19,Blanton+17,Bundy+15a,Belfiore+19,Westfall+19,Cherinka+19,Wake+17,Law+15,Law+16,Law+21,Yan+16,Yan+16a,Drory+15,Gunn+06,Smee+13,riffel}. The AGN luminosities and stellar masses for $224$ AGN host galaxies from this galaxy sample are taken from the {\sc megacubes}\footnote{avaliable at: manga.if.ufrgs.br or manga.linea.org.br} presented by \cite{riffel}. These AGN luminosities are based on the MaNGA DR12 integrated  $[\ion{O} {III}]$$\lambda5007\textup{~\AA}$ luminosities using an aperture around the galactic centre \citep{riffel,Riffel+21}. Our AGN host galaxies sample from MaNGA is significantly reduced due to the requirements of AGN activity and kinematic position angles with the required uncertainties (see  \citealt{sandra} for more details on the kinematic position angles).

For the following analysis on the morphology trend, we used the MaNGA PyMorph photometric and deep-learning morphological catalogue from \cite{sanchez} to classify the galaxies as early-type (ETG) or late-type (LTG). We used two variables from the catalogue for the classification: the T-type $\leq 0$ for ETGs and T-type $>0$ for LTGs, as well as the probability of a galaxy being an LTG ($P_{LTG}$) being $<0.1$ for ETGs and $>0.9$ for LTGs. We used this classification to minimise contamination between ETGs and LTGs as that is a major part of our analysis. For the purpose of the morphology analysis, we then used a MaNGA galaxy sub-sample of 143 galaxies with well determined early- or late-type classifications.

\subsection{Stellar mass range of SAMI vs MaNGA}
\begin{figure}
	\includegraphics[width=\columnwidth]{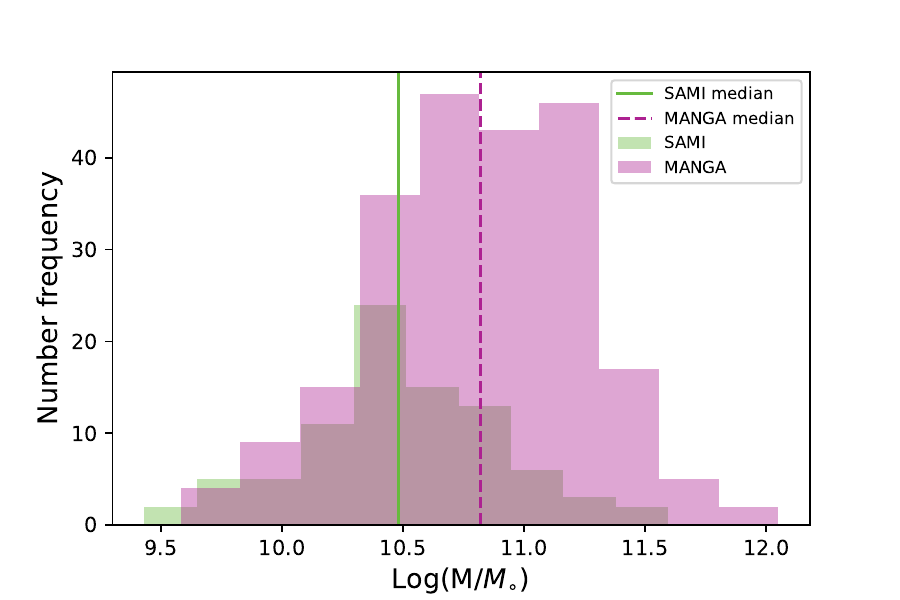}
    \caption{Histogram showing the stellar mass distributions of the SAMI (green) and MaNGA (pink) galaxy samples, with their stellar mass medians (green solid line, pink dashed line, respectively). The mass distributions investigated by the two samples are different as MaNGA probes more higher mass galaxies.}
    \label{fig:masses_MANGA_SAMI}
\end{figure}

The galaxies in MaNGA extend to higher redshifts than SAMI, and are more massive. Both distributions have varied morphological mix, with a predominance of S0s and spirals. Figure \ref{fig:masses_MANGA_SAMI} shows the stellar mass distributions of galaxies from SAMI \citep{SAMI_swoi} in green, and from MaNGA \citep{rembold} in pink. The MaNGA sample has a much higher fraction of higher mass galaxies than SAMI, and as a result the two distributions only resemble each other for lower mass galaxies. Their stellar mass medians are significantly different: $10^{10.48}M_{\odot}$ (SAMI) and $10^{10.86}M_{\odot}$ (MaNGA). This will be relevant for the discussion of our results.

\section{Methods} \label{methods}
In this section we present the methods by which we extracted kinematic misalignment and AGN luminosities for SAMI and MaNGA galaxies.
\subsection{Kinematic misalignment angles}
\label{misalignment_angle}
\begin{figure*}
    \centering
    \begin{subfigure}{0.49\textwidth}
        \centering
        \includegraphics[width=\columnwidth]{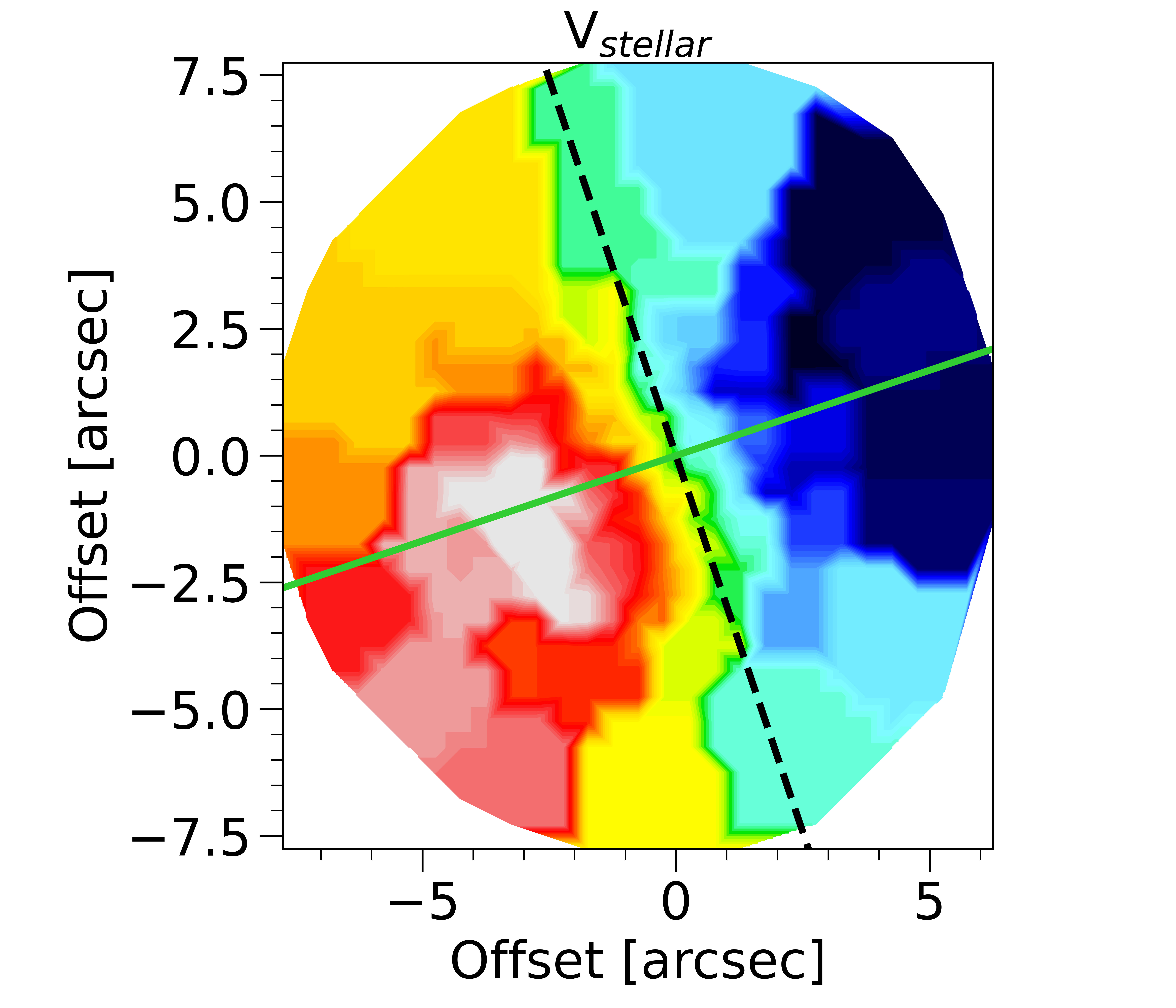}
        \caption{Velocity map of the stellar component.}
        \label{fig:a}
    \end{subfigure}
    \begin{subfigure}{0.49\textwidth}
        \centering
        \includegraphics[width=\columnwidth]{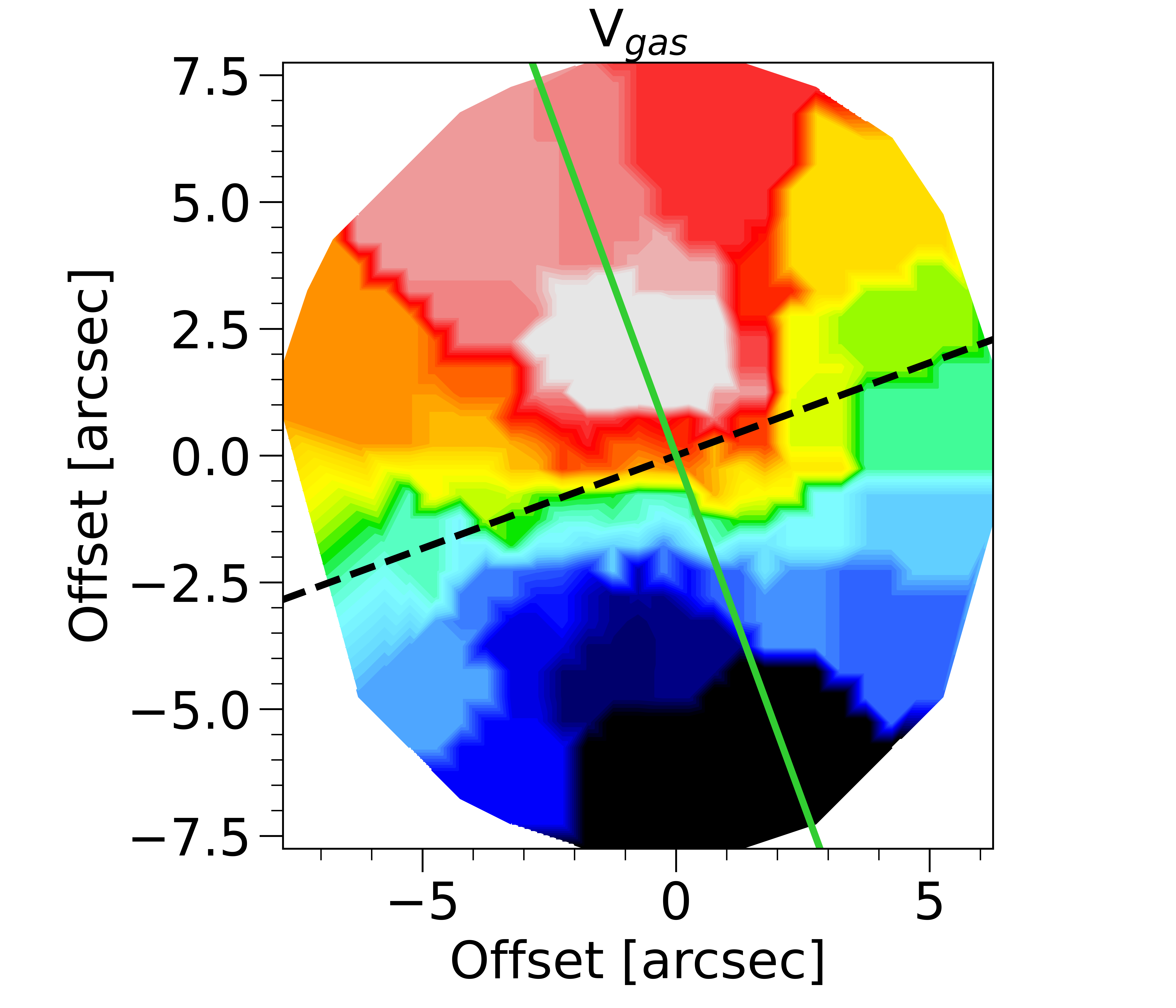}
        \caption{Velocity map of the gas component.}
        \label{fig:b}
    \end{subfigure}
    \caption{Velocity maps for one of the galaxies from our sample (SAMI Galaxy ID 323593). On the left is the stellar velocity distribution, and on the right is the gas velocity distribution. The green lines indicate the best-fit of the kinematic position angle (PA), determined as $PA_{stars}=289 \pm 3 ^{\circ}$, and $PA_{gas}=200 \pm 1 ^{\circ}$. The PA orientation is measured from the North (where $PA=0$) towards East (i.e., top to the left of the figure). This results in the galaxy misalignment, $\Delta PA = 89 \pm 3 ^{\circ}$. The dashed lines indicate the zero-velocity lines, used as reference for the kinematic position angles. The red colour indicates redshift (receding motion), and blue colour indicates blueshift (approaching motion). By considering the receding and approaching parts of the velocity maps, one can visually see the offset between stellar and gas motions in the galaxy.}
    \label{fig:velocity_maps}
\end{figure*}

The kinematic position angle of stars and gas in a galaxy is defined along the kinematic major axis (i.e., measured from the North towards the East, to the highest velocities in the approaching side of the rotation, which corresponds to blueshifted velocities on a velocity map); it is found through a model discussed in more detail in \cite{sandra}. The kinematic position angles are found by applying this model to the 2D stellar and gas velocity maps and represent the direction of rotation for gas and stars. Figure \ref{fig:velocity_maps} shows a velocity map of the stellar and gas components of one of the host galaxies we analysed (SAMI Galaxy ID 323593). The red colour on the velocity maps corresponds to redshifted velocities, and the blue to blueshifted velocities. The gas and stellar components in that galaxy are misaligned by around $90^{\circ}$, which can be visually seen by the offset between the position of the redshifted stellar and gas components of the galaxy.

For SAMI we use the kinematic misalignment angles as published by \cite{sandra}, and for MaNGA we use the velocity maps from {\sc megacubes}\footnote{avaliable at: manga.if.ufrgs.br or manga.linea.org.br}  \citep{riffel}, as analysed by Raimundo et al. 2024, submitted. We consider misalignment to be the angle between the kinematic position angle of stars vs that of gas in a galaxy. Equation \ref{eq:misalignment} shows how the kinematic misalignment angle, $\Delta PA$, is found using the kinematic position angles, $PA_{\mathrm{stars}}$ and $PA_{\mathrm{gas}}$, of stars and gas (respectively) in a galaxy:

\begin{equation} 
    \Delta PA = | PA_{\mathrm{stars}} - PA_{\mathrm{gas}} |.
	\label{eq:misalignment}
\end{equation}

In this work we consider galaxies to be aligned for $0\degree \leq \Delta PA < 45\degree$ and misaligned for $45\degree \leq \Delta PA \leq 180\degree$ (although some research considers $30\degree \leq \Delta PA \leq 180\degree$; e.g. \citealt{tim_davis}). The reason for the cut of $45\degree$ is to ensure the misalignment present came from an external process such as a merger, as opposed to small-scale misalignment caused by for example bars in galaxies (which host non-circular orbits because the potential is no longer axisymmetric, which would be naively interpreted as small misalignment; \citealt{tim_davis, sandra}). We will discuss in Section \ref{results}, and show quantitatively in Figure \ref{fig:early_vs_late_oxygen}, that our results are unaltered when assuming a less stringent cut of $30\degree$, as adopted by other groups (e.g., \citealt{tim_davis}, \citealt{lagos}), which is expected as not many sources in our sample are located in the  $30\degree$ to $45\degree$ misalignment range.
Therefore, in this paper, we test the hypothesis that all misalignment angles larger than $45\degree$ originate from externally acquired gas, as in \cite{sandra}. Our working hypothesis does not rule out the possibility that even aligned gas in ETGs is of external origin, as suggested by several authors (e.g., \citealt{bertola}) and simulations \citep{lagos}. Possibly a high (but still unknown) fraction of the gas could be of external origin, because the gas relaxes from misaligned configurations in short timescales \citep{max_baker}.

\subsection{Determination of AGN luminosities} \label{AGN_lums}
All of the galaxies used in our analysis have confirmed AGN activity (see Sections \ref{imaj_quality_cut} and \ref{manga_sample}). To obtain the AGN luminosities for the galaxies in the SAMI sample, we integrated the  $[\ion{O} {III}]$($\lambda5007\textup{~\AA}$) fluxes in pixels inside a nuclear aperture using Python's \texttt{photutils} package. 

For the SAMI sample we assumed a radius of aperture equal to double the full width half maximum (FWHM) of the point spread function of the observations for each galaxy. Light from a singular object (such as unresolved emission from the AGN) can spread out over many pixels due to dispersion caused by atmospheric effects, and considering the double of the FWHM makes it more likely to collect all of the emission. The  $[\ion{O} {III}]$ flux values of all pixels inside the aperture can therefore be summed to find an approximate total AGN flux, as  $[\ion{O} {III}]$($\lambda5007\textup{~\AA}$) flux is commonly used as a proxy for total AGN flux \citep{lamastra}. Since there is extended ionised gas in every galaxy, these fluxes could be contaminated with galaxy emission. To test this, we used the same aperture radius ($1.5$ kpc) for all SAMI galaxies to investigate the changes in  $[\ion{O} {III}]$ flux; we found little difference of about $20\%$, and henceforth in our analysis assume there is little contamination to the  $[\ion{O} {III}]$ fluxes.

Apertures centred at the galactic centre of a galaxy's flux map are traditionally used for obtaining an estimate of AGN flux. However, its limitation is the assumption that most AGN emission comes from the centre of a galaxy, which may not always be the case. We therefore tested the pixels around the galactic centre for AGN emission using Baldwin, Phillips \& Terlevich (BPT) diagrams \citep{BPT}, and concluded that for the SAMI galaxy sample, a nuclear aperture is a very good approximation, since the flux integration within the aperture provides a similar flux value to that determined when integrating all the pixels with AGN excitation throughout the galaxy. Additionally, this allowed for a more direct comparison with MaNGA AGN luminosities, which were also found using the  $[\ion{O} {III}]$($\lambda5007\textup{~\AA}$) flux integrated in a nuclear aperture in the same way \citep{riffel}. 

The luminosity from  $[\ion{O} {III}]$($\lambda5007)$ transition is in general a good approximation for AGN bolometric luminosity \citep{lamastra}, because it arises from gas in the narrow line region that was ionised due to AGN radiation.  $[\ion{O} {III}]$ is typically the most prominent emission line in an AGN spectrum and does not suffer significant extinction from the dusty torus, making it a more reliable AGN luminosity tracer than other emission lines. We converted  $[\ion{O} {III}]$ fluxes to  $[\ion{O} {III}]$ luminosities using the cosmology stated in Section \ref{Intro}.

We tested the effect of dust on the  $[\ion{O} {III}]$ luminosity for the SAMI galaxies for bolometric luminosities of the AGN in our sample. By measuring the Balmer decrement, which traces the dust abundance in a galaxy, we found a dust correction factor for the  $[\ion{O} {III}]$ luminosity, $L_{ [\ion{O} {III}]}$, using \citealt{bassani}. We found that the dust correction for most of the SAMI galaxies of interest to our work was minor, of the order of $1\%$ to $2\%$. In our analysis, we thus disregard any dust correction and use the observed  $[\ion{O} {III}]$ luminosities; this also allowed us to make a closer comparison with the MaNGA  $[\ion{O} {III}]$ luminosities which are uncorrected for dust extinction. 

The dust corrected  $[\ion{O} {III}]$ luminosities ($L_{ [\ion{O} {III}]}^C$) were only used to compute approximate bolometric luminosities of our sources, using
\begin{equation}
    L_{\rm bol} = L_{ [\ion{O} {III}]}^C \times BC.
	\label{eq:bolometric_lums}
\end{equation}
where $BC$ is the bolometric correction, taken as $87$ in extinction corrected luminosity range of $10^{38}$ to $10^{40}$  erg\ s$^{-1}$, and $142$ in extinction corrected luminosity range of $10^{41}$ to $10^{42}$  erg\ s$^{-1}$ \citep{lamastra}. Our AGN L$_{ [\ion{O} {III}]}$ correspond to bolometric luminosities in the range L$_{\rm bol}$ = $10^{40}$ to $10^{43}$ erg\ s$^{-1}$, which are well below the bolometric luminosities measured for quasars in the same redshift range from the Sloan Digital Sky Survey (SDSS) Data Release 16 \citep{wu} ($10^{44}$ to $10^{47}$ erg\ s$^{-1}$). They estimated the bolometric luminosity using the continuum luminosity (excluding Fe II emission) at 3 rest-frame wavelengths dependent on the AGN SEDs \citep{wu}. This indicates that the AGN in our sample are low to median luminosity AGN. Our measured low bolometric luminosities also support the findings of \cite{ilha19} and \cite{sandra} that the AGN are not driving the observed kinematic misalignment angles by means of gas outflows (as for example occurs in a small number of 'Red Geyser' sources \citep{ilha}). 

\section{Results} \label{results}
Here we summarise the results of our investigation of AGN luminosity as a function of misalignment angle for the combined sample of SAMI and MaNGA AGN host galaxies. 

\begin{figure*}
	\includegraphics[width=14cm]{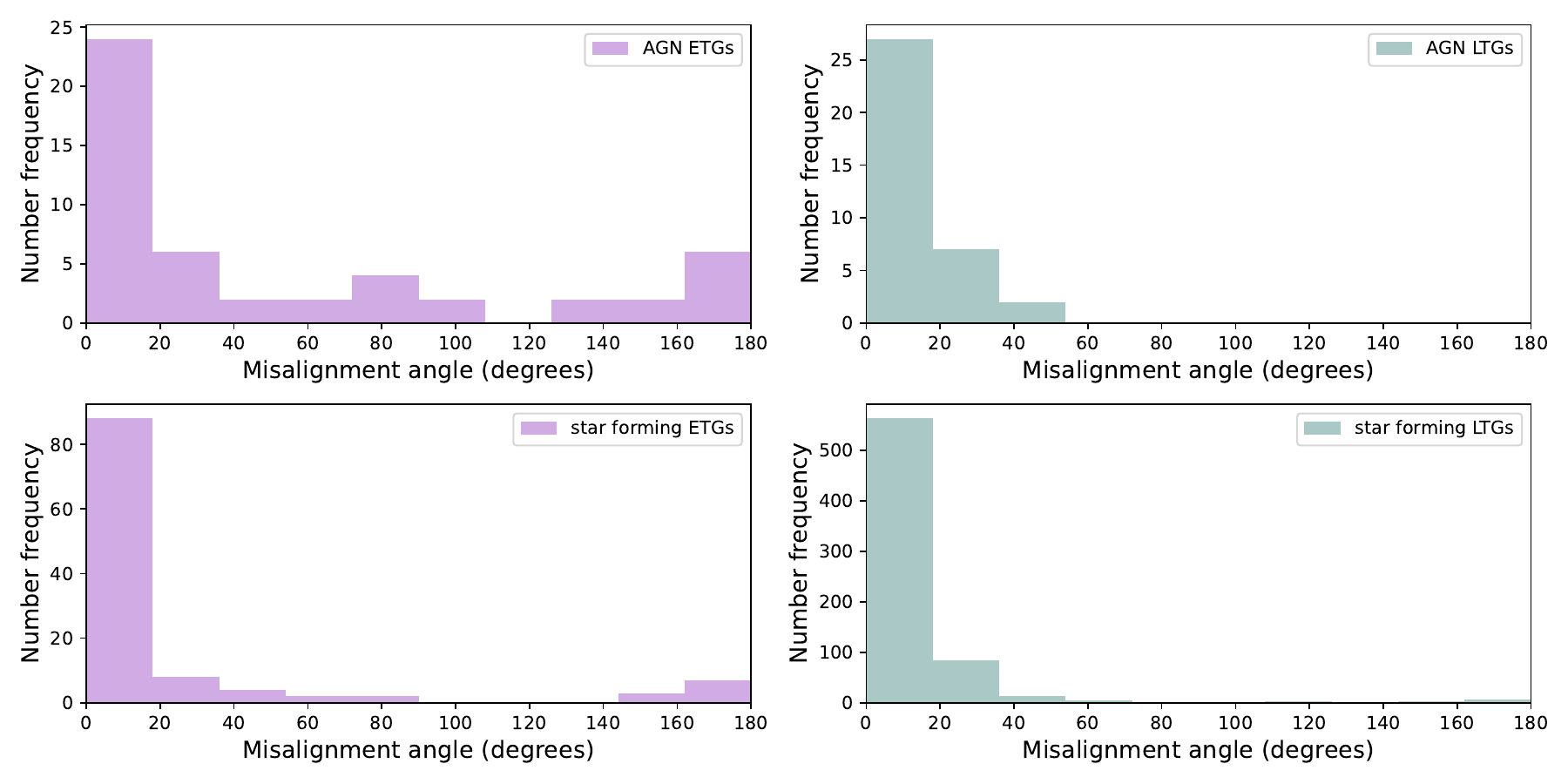}
    \caption{Number frequency of AGN (top) and star-forming galaxies (bottom) from SAMI as a function of misalignment angle, separated between ETGs (pink) and LTGs (green). It becomes apparent AGN ETGs are the most likely to be misaligned out of the 4 groups, particularly as AGN galaxies are a much smaller sample size compared to star forming galaxies.}
    \label{fig:star_agn_hist}
\end{figure*}

\subsection{AGN luminosity as a function of misalignment angle for SAMI and MaNGA galaxies} \label{main_sami_result}
\begin{figure*}
	\includegraphics[width=14cm]{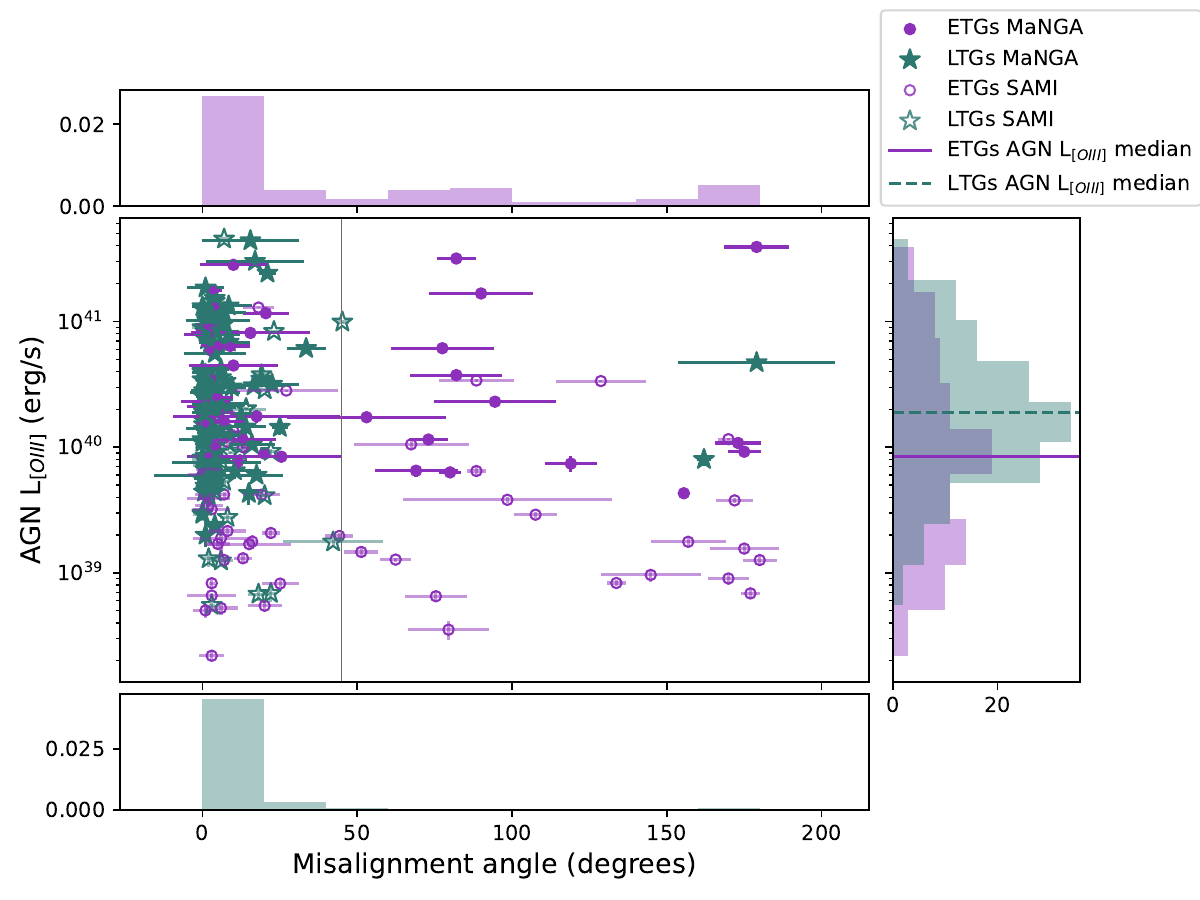}
    \caption{AGN $L_{ [\ion{O} {III}]}$ as a function of misalignment angle ($\Delta PA = |PA_{stellar}-PA_{gas}|$) for SAMI and MaNGA AGN host galaxies. Each point on the plot is an ETG host (purple circle) or an LTG host (green star). The grey line indicates the misalignment ($\Delta PA $) of $45^\circ$. SAMI host galaxies are represented by hollow markers. Shown are also histograms with the AGN $L_{ [\ion{O} {III}]}$ and normalised misalignment angle distributions between ETG hosts (top) and LTG hosts (bottom). ETG hosts and LTG hosts display different behaviour; LTG hosts are mainly aligned. ETG hosts have the full misalignment range, and their mean $L_{ [\ion{O} {III}]}$ seems unaffected by the misalignment angle. ETG hosts also have a higher fraction of lower luminosity AGN compared to LTG hosts. Both ETG hosts and LTG hosts have a similar $L_{ [\ion{O} {III}]}$ range.}
    \label{fig:early_vs_late_oxygen}
\end{figure*}

Figure \ref{fig:early_vs_late_oxygen} shows the L$_{ [\ion{O} {III}]}$ AGN luminosity as a function of kinematic misalignment for SAMI and MaNGA host galaxies, where each data point corresponds to a single host galaxy. SAMI host galaxies are represented by hollow points. An important part of our analysis was the galaxy grouping into ETG hosts (purple circles) and LTG hosts (green stars) to investigate the effects of fresh gas on different galaxy types. This is because misalignment is more commonly found in ETGs (e.g., \citealt{bryant_croom}, \citealt{ristea}, \citealt{sandra}), and it has been suggested that the AGN in S0 galaxies are fuelled by externally accreted gas \citep{davies_maciejewski}. In general there is a difference in the misalignment angle distributions between the AGN and star forming (non-AGN) populations of SAMI galaxies. Figure \ref{fig:star_agn_hist} clearly shows that there is a higher number frequency of AGN galaxies with misalignment compared to the star-forming galaxies (particularly for ETGs), similar to what was seen in \citealt{sandra}. Shown are also histograms with AGN luminosity and kinematic misalignment distributions for ETG hosts and LTG hosts. There appears to be a clear separation between ETG hosts and LTG hosts' AGN luminosities and misalignment angle relations, present in both SAMI and MaNGA host galaxy samples. Our most important finding is that ETG hosts seem to have an approximately constant AGN luminosity range irrespective of the misalignment angle. To test this, we conducted a Mann-Whitney U test on the distribution of aligned and misaligned ETGs in both SAMI and MaNGA populations. It is a statistical test to determine whether two samples come from the same parent distribution. The null hypothesis is that the parent distribution is the same for both. The p-value measures the probability of obtaining the observed difference between distributions, assuming the null hypothesis is true. If the p-value is below or equal to the significance level, the difference between two samples is statistically significant. We take the significance level to be 0.05 for all Mann-Whitney U tests conducted in this paper. We tested the AGN luminosity distributions of aligned and misaligned ETGs from our sample. Our p-value for this test is $0.322$, meaning there is no significant statistical difference between aligned and misaligned ETGs from our galaxy sample.

The top and bottom histograms of Figure \ref{fig:early_vs_late_oxygen} show the distribution of kinematic misalignment angles for ETG and LTG hosts, respectively. ETG hosts show misalignment angles that span the full possible kinematic misalignment angle range [$0-180^\circ$], but most (56 out of 89) are aligned. This, however, does not mean that the aligned ETGs do not have external gas in them - they could have been misaligned in the past, or happened to be aligned after an external event. Most ETGs are consistent with having externally accreted gas (\citealt{tim_davis}, \citealt{raimundo17}). MaNGA contains more galaxies than SAMI, including more misaligned LTG hosts, which is advantageous for statistical analysis. Still, almost all LTG hosts are aligned (only 3 out of 140 have $\Delta PA \geq 45^\circ$). Misaligned LTG hosts from MaNGA appear to be able to have any misalignment angle, similar to what is seen for ETG hosts. 
\\
The histogram on the right of Figure \ref{fig:early_vs_late_oxygen} shows the AGN L$_{ [\ion{O} {III}]}$ distribution, with the medians for ETG hosts and LTG hosts indicated as solid and dotted lines, respectively. The AGN L$_{ [\ion{O} {III}]}$ distributions are similar; LTG hosts show an approximately Gaussian distribution, while ETG hosts have two peaks as they have a higher fraction of lower AGN luminosity galaxies in their distribution. This is expected, as ETGs tend to have broad Eddington ratio distributions \citep{mcconnell}.  The AGN luminosity range of ETG hosts is very comparable to that of LTG hosts. 

\section{Discussion} \label{discussion}
\subsection{AGN luminosity as a function of misalignment angle}
We investigated the relation between AGN luminosity and misalignment angle of AGN host galaxies. Clear differences between ETG and LTG hosts emerged in terms of both AGN luminosities and misalignment angles for AGN host galaxies from both SAMI and MaNGA (Figure \ref{fig:early_vs_late_oxygen}). Our kinematic misalignment results agree with previous findings for galaxies without AGN (e.g., \citealt{pizzella}, \citealt{tim_davis}), and therefore indicate that the distribution of kinematic misalignment does not show a substantial difference between AGN host galaxies and galaxies without AGN (see also  \citealt{sandra}). The new result we present is that the AGN luminosity does not correlate with kinematic misalignment in our combined (SAMI+MaNGA) sample.

This is in line with previous findings by \cite{ilha19} that did not find evidence for systematically larger misalignment angles in luminous AGN, using a smaller sample of MaNGA AGN hosts. To summarise the results for AGN host galaxies from SAMI and MaNGA, LTG hosts are mainly presenting kinematically aligned stellar and gas rotation. ETG hosts are more likely to show kinematic misalignment, which appears to have little to no effect on their AGN luminosities. Although this could suggest no clear trend between AGN luminosity and misalignment angle, it is apparent that the luminosity distribution versus misalignment angle differs between ETG hosts and LTG hosts. Since all are AGN host galaxies, this implies ETG hosts and LTG hosts could have different ways of fuelling their AGN. 

\begin{figure}
	\includegraphics[width=\columnwidth]{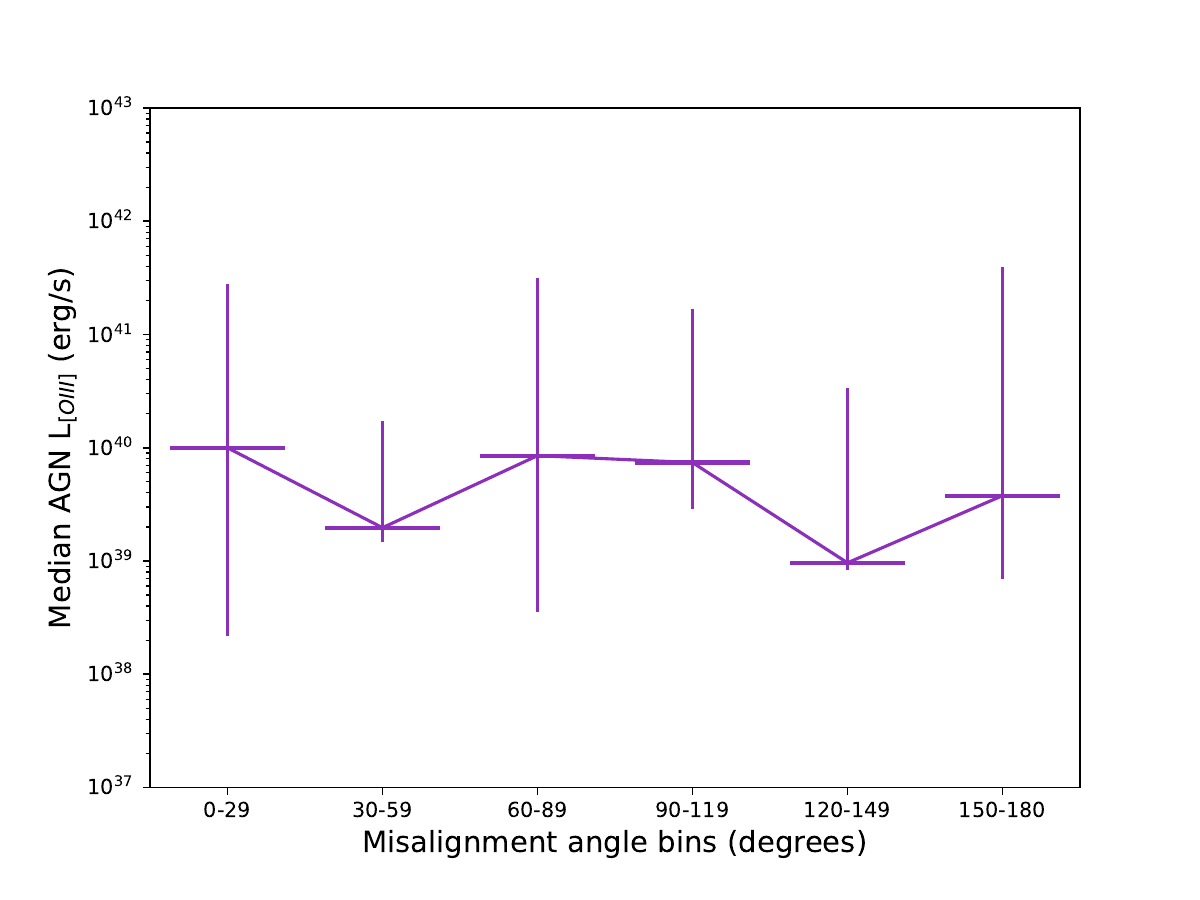}
    \caption{Median AGN $L_{ [\ion{O} {III}]}$ for kinematic misalignment angle bins for ETG hosts from SAMI and MaNGA. The maximum and minimum AGN $L_{[\ion{O} {III}]}$ values for each misalignment bin are represented as the error bars.}
    \label{fig:etg_luminosity}
\end{figure}

Our main result is that the AGN luminosity range as a function of the misalignment angle stays approximately constant, as illustrated in Figure \ref{fig:etg_luminosity} which reports the median AGN $L_{ [\ion{O} {III}]}$ from ETG hosts from SAMI and MaNGA as a function of misalignment, in bins of equal size.  The high fraction of misaligned galaxies with an AGN could indicate the two are linked. Hence, a question remains if the AGN fuelling process is equally efficient for aligned and misaligned ETG hosts. ETG hosts often have a lower amount of native gas than LTG hosts (e.g., \citealt{gallagher}, \citealt{Lees}, \citealt{Young}), and could acquire significant fresh gas through a large-scale external process that often leads to misalignment. Such an external event is also one of the best candidate mechanisms to allow ETGs to obtain enough gas to fuel their AGN at later times. A recent study also showed that AGN host galaxies and control galaxies from the LLAMA survey have in principle the same amount of 'internal' cold gas \citep[e.g.,][]{Rosario}. In simulations, such as that of \citet{lagos}, ETGs can end up misaligned or aligned after external gas accretion event dependent on their cold gas content before the event. Additionally, \citealt{Rembold+24} studied the environment of AGN hosts from MaNGA, and found that at least a part of them has been triggered by tidal interactions with nearby-galaxies.

External gas accretion can still result in aligned stellar to gas kinematics if, for example, the two galaxies undergoing a merger are on the same kinematic axis of rotation. This means aligned ETG hosts could also have their AGN fuelling be due to external interactions, but happened to be aligned, or already relaxed to kinematic alignment (e.g. \citealt{vandevoort15}, \citealt{raimundo21}) after their initial misalignment (for example initial $\Delta$PA$<90^{\circ}$). Our result for ETG hosts could therefore be indicative of an AGN fuelling model in which the misalignment provides the fresh gas as fuel, but it is another factor, for example the total gas mass, that determines AGN luminosity. Finding a correlation between AGN activity and misalignment is notoriously difficult due to their different timescales - the flickering timescale of AGN activity of  $\sim 10^5$ yrs  \citep{schawinski2015activegalacticnucleiflicker} is significantly shorter than the lifetime for misalignment of  $\sim 10^8$ yrs (e.g., \citealt{taylor}, \citealt{choi}). This is why we first selected the longer-lived physical feature (kinematic misalignment), and then investigated the AGN activity within that sample in an approach similar to \citealt{sandra}.

If ETGs and LTGs have different AGN fuelling mechanisms, their AGN luminosity distributions could reflect that. The AGN luminosity distributions for ETG hosts and LTG hosts are reminiscent of each other, although their AGN luminosity medians are quite different. To evaluate whether the two samples (ETG hosts and LTG hosts) come from the same parent population, we conducted a Mann-Whitney U test on their AGN luminosity distributions. Our p-value in this test is $0.0001$, indicating there is strong evidence to reject the null hypothesis, meaning there is a significant difference between ETGs and LTGs AGN luminosity distributions. In other words, we cannot say that ETG hosts and LTG hosts come from the same parent population. All galaxies considered are AGN host galaxies and had a similar survey selection approach, so this statistical difference between the ETG and LTG hosts' AGN luminosity distributions could support that the AGN fuelling processes for ETGs and LTGs are different. However, some recent work in the field has shown that some fuelling processes may be common, since galaxies can become reactivated in their star formation rate and AGN activity via minor mergers \citep{navarro}, which could suggest that even if most of the AGN hosts are aligned, their SMBH may still have a fuelling contribution from external interactions. 

\subsubsection{Eddington ratios distributions}

\begin{figure*}
	\includegraphics[width=14cm]{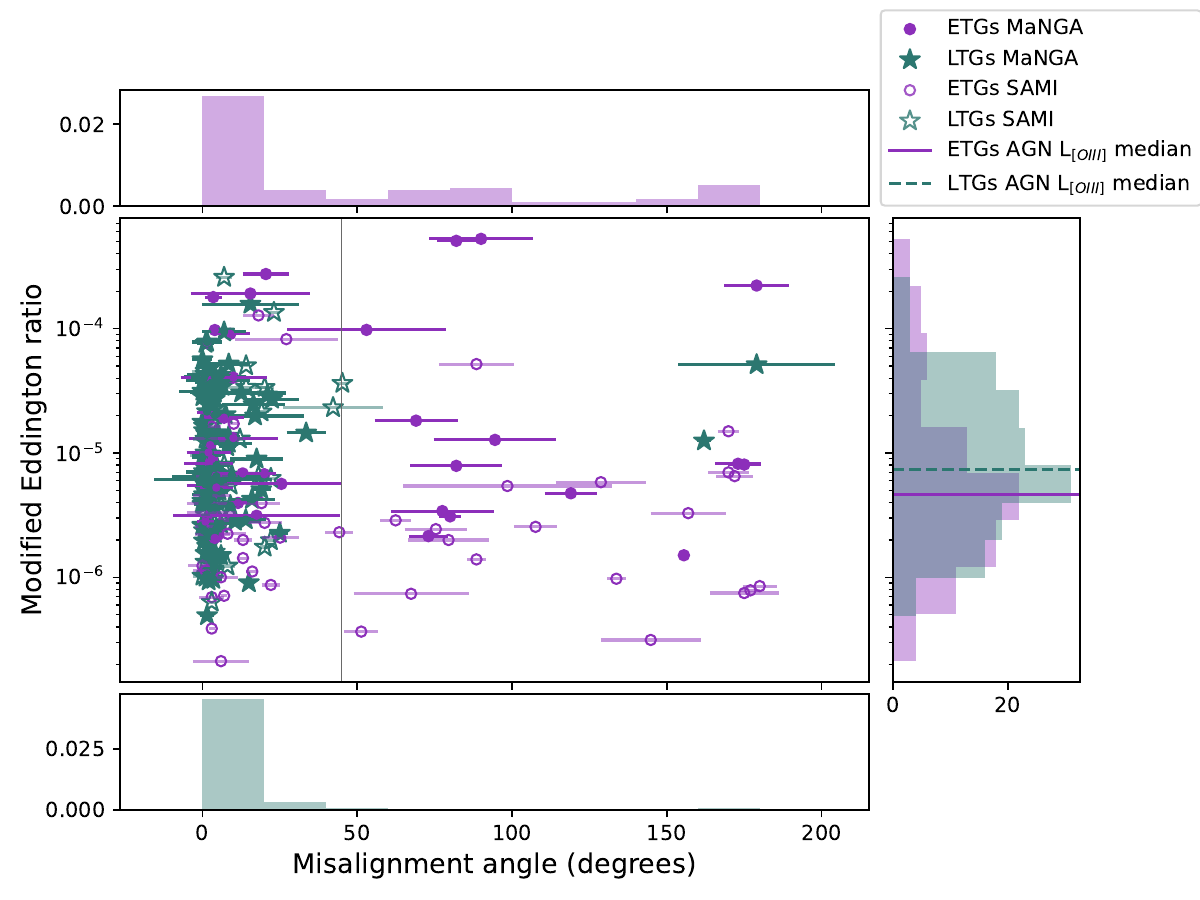}
    \caption{Eddington ratio as a function of misalignment angle ($\Delta PA = |PA_{stellar}-PA_{gas}|$) for SAMI and MaNGA AGN host galaxies. Each point on the plot is an ETG host (purple circle) or an LTG host (green star). The grey line indicates the misalignment ($\Delta PA $) of $45^\circ$. SAMI host galaxies are represented by hollow markers. Shown are also histograms with the Eddington ratio and normalised misalignment angle distributions between ETG hosts (top) and LTG hosts (bottom). The same general trend is seen as in Figure \ref{fig:early_vs_late_oxygen}; LTG hosts are mainly aligned. ETG hosts have the full misalignment range, and their mean $L_{ [\ion{O} {III}]}$ seems unaffected by the misalignment angle. }
    \label{fig:eddington_ratios_result}
\end{figure*}

The differences between SMBH masses of ETGs and LTGs results in different expected Eddington ratios  between them \citep{mcconnell}, which could have an impact on our findings. Figure \ref{fig:eddington_ratios_result} shows the Modified Eddington ratios of ETG and LTG hosts as a function of kinematic misalignment for SAMI and MaNGA samples. We used the stellar mass and black hole mass relation from \citep{reines} (see their Equations 4 and 5) to calculate the black hole masses for our sample. We also used $L_{ [\ion{O} {III}]}$ (instead of the bolometric luminosities), and hence found Modified Eddington ratios. The general trend seen in Figure \ref{fig:early_vs_late_oxygen} between ETGs and LTGs is present once more, implying the different SMBH masses of ETG and LTG hosts did not have a significant impact on our analysis.

\subsection{The effect of stellar mass}
\begin{figure*}
	\includegraphics[width=15cm]{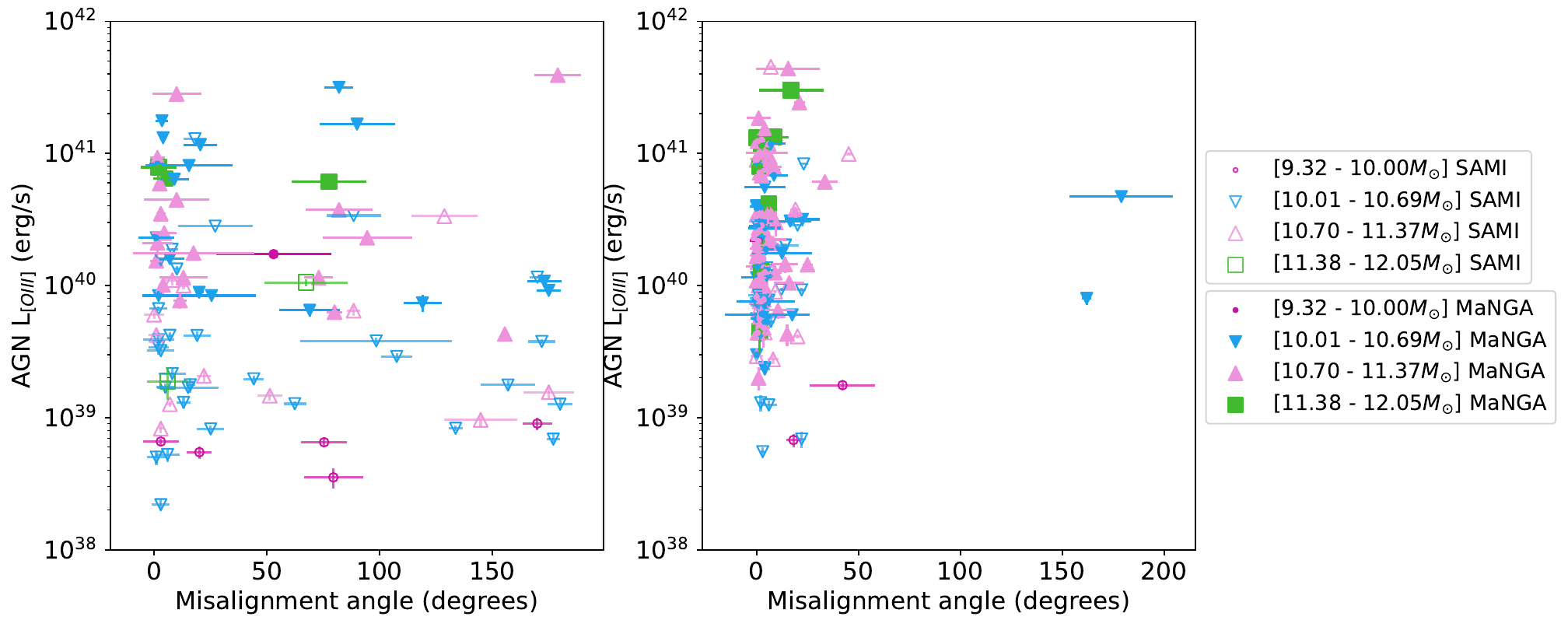}
    \caption{AGN $L_{ [\ion{O} {III}]}$ as a function of misalignment angle ($\Delta PA = |PA_{stellar}-PA_{gas}|$) for ETG hosts (left) and LTG hosts (right). The y-axis is in logarithmic scale. Each point represents an AGN host galaxy from SAMI or MaNGA, colour-coded between 4 stellar mass bins, with the stellar mass ranges in each bin indicated in the legend on the plot. SAMI AGN host galaxies are represented by hollow markers. No stellar mass dependence emerges.}
    \label{fig:manga_sami_mass}
\end{figure*}

It is worth mentioning that there are some differences between the host galaxies investigated from SAMI and MaNGA, such as the different stellar masses probed (Section \ref{manga_sample}) or MaNGA's larger sample size. In SAMI, there is a much larger number of AGN with lower luminosity misaligned in ETG hosts compared to MaNGA (which may explain why there is the second distribution peak for ETG hosts in the AGN luminosity histogram in Figure \ref{fig:early_vs_late_oxygen}). This could be because SAMI AGN where selected including radio, X-ray and infrared detected AGN which may not be as strong in  $[\ion{O} {III}]$ luminosity as the MaNGA AGN (selected using BPT and WHAM diagrams). Another difference is how the misaligned ETG hosts from MaNGA reach the highest AGN luminosities, which was not observed in the SAMI sample.  Figure \ref{fig:manga_sami_mass} shows the AGN luminosity and misalignment angle dependence plot. Each point represents a host galaxy, colour-coded according to 4 stellar mass logarithmic bins of equal size. There appears to be no dependence between stellar mass and kinematic misalignment angle for ETG hosts or LTG hosts, other than there are more higher stellar mass host galaxies in MaNGA. The (very few) misaligned LTG hosts are in the second to lowest mass bin. This could suggest those LTGs are misaligned because of their lower masses, which mean their gas reservoirs are easier to perturb with external interactions, or that since misaligned LTGs are a rarer occurrence, they appear in the MaNGA sample which contains a larger number of galaxies.

As expected, most of the LTG hosts are not misaligned ($\Delta$PA $<45^{\circ}$), likely because they have a considerable reservoir of native gas already. The gas and stars align faster due to dissipation between the incoming gas and the native gas (\citealt{kannappan01}, \citealt{bassett17}). We are therefore less likely to see misaligned LTG hosts due to their smaller abundance and the short amount of time they are misaligned for. Due to a high number of LTG hosts that are aligned, it is likely that  misalignment did not trigger their AGN fuelling and that most of their AGN are fuelled via other (secular) mechanisms. The gas in LTG hosts may already be fuelling the AGN or they may have larger fractions of recycled stellar gas from intermediate age stars (\citealt{Riffel+24}, \citealt{choi}). Additionally, aligned LTG hosts have a higher AGN luminosity range than misaligned LTG hosts. This behaviour is not seen with ETGs, which could once again imply a different AGN fuelling process between ETG hosts and LTG hosts.

\section{Conclusions}
The main processes triggering the fuelling of Active Galactic Nuclei (AGN) remain unknown. The main aim of this paper was to investigate the effects, if any, of stellar to gas kinematic misalignment ($ \Delta PA \geq 45^\circ$) on Active Galactic Nuclei luminosity for early- and late-type galaxies hosting known AGN. We used SAMI and MaNGA samples of low redshift galaxies (86 in $0.004<z<0.095$ redshift range and 143 galaxies in  $0.01<z<0.15$ redshift range, respectively) to investigate if galaxies with acquired gas have AGN that are more luminous. The dust corrected AGN  $[\ion{O} {III}]$ luminosities of the host galaxies investigated correspond to bolometric luminosities in the range $10^{40}$ to $10^{43}$erg\ s$^{-1}$, which indicate that the AGN in our sample are low to moderate luminosity AGN. This luminosity range supports the kinematic misalignment being caused by an external process and not driven by AGN outflows.  

We studied the AGN luminosity as a function of kinematic misalignment between the stellar and gas components in the galaxies, grouped for ETG hosts and LTG hosts as well as different stellar masses of the host galaxies. Our results of ETG and LTG hosts agree with previous findings that there is a clear difference between ETG and LTGs when it comes to kinematic misalignment. In both samples, LTG hosts are mainly kinematically aligned. ETG hosts show misalignment angles that span the full range of stellar to gas kinematic angles $\Delta PA $[$0-180^\circ$], and they have the same AGN luminosity range irrespective of misalignment angle. The AGN luminosity ranges of ETG hosts and LTG hosts are very comparable. Our results, combined with previous work - higher observed fraction of AGN in galaxies with misalignment \citep{sandra} - suggest that the AGN fuelling mechanisms differ between ETG hosts and LTG hosts in our sample. We find that there is no trend between $\Delta PA$ and AGN luminosity in ETGs. This is consistent with the expectation that if a significant fraction of ETGs acquire gas externally \citep{bertola}, both aligned and misaligned ETG hosts may have their AGN fuelling gas originating in the same process. This means they were likely fuelled by an external gas accretion event, but happened to end up kinematically aligned. More spatially-resolved observations of interacting LTG hosts could help identifying other mechanisms by which their AGN could be fuelled.

\section*{Data Availability}
The MaNGA MEGACUBES used in this work are publicly available through a web interface at https://manga.linea.org.br/ and https://manga.if.ufrgs.br/. The SAMI data are publicly available at https://datacentral.org.au/.

\section*{Acknowledgements}
The authors would like to thank the anonymous referee for a timely and constructive referee report which improved the clarity of the work. MW would like to extend her thanks to the Astronomy department at the University of Southampton for their incredible and continued support over the academic year 2023/2024. MW would also like to extend her thanks to Max Baker for the extremely valuable discussions about kinematically misaligned galaxies. This work was supported by the Science and Technology Facilities Council (STFC) of the UK Research and Innovation via grant reference ST/Y002644/1 and by the European Union's Horizon 2020 research and innovation programme under the Marie Sklodowska-Curie grant agreement No 891744 (SIR).
This research has made use of MaNGA data from SDSS IV. Funding for the Sloan Digital Sky Survey IV has been provided by the Alfred P. Sloan Foundation, the U.S. Department of Energy Office of Science, and the Participating Institutions. SDSS acknowledges support and resources from the Center for High-Performance Computing at the University of Utah. The SDSS web site is www.sdss4.org. FS acknowledges partial support from the European Union’s Horizon 2020 research
and innovation programme under the Marie Skłodowska-Curie grant agreement No 860744 (grant coordinator F. Shankar).
RR acknowledges support from  Conselho Nacional de Desenvolvimento Cient\'{i}fico e Tecnol\'ogico  ( CNPq, Proj. 311223/2020-6,  304927/2017-1, 400352/2016-8, and  404238/2021-1), Funda\c{c}\~ao de amparo \`{a} pesquisa do Rio Grande do Sul (FAPERGS, Proj. 19/1750-2 and 24/2551-0001282-6) and Coordena\c{c}\~ao de Aperfei\c{c}oamento de Pessoal de N\'{i}vel Superior (CAPES, Proj. 0001).

This research made use of Astropy, \href{http://www.astropy.org}{http://www.astropy.org} a community-developed core Python package for Astronomy \cite{astropy:2013}. 



\bibliographystyle{mnras}
\bibliography{main_text} 

\begin{thebibliography}{}
\makeatletter
\relax
\def\mn@urlcharsother{\let\do\@makeother \do\$\do\&\do\#\do\^\do\_\do\%\do\~}
\def\mn@doi{\begingroup\mn@urlcharsother \@ifnextchar [ {\mn@doi@} {\mn@doi@[]}}
\def\mn@doi@[#1]#2{\def\@tempa{#1}\ifx\@tempa\@empty \href {http://dx.doi.org/#2} {doi:#2}\else \href {http://dx.doi.org/#2} {#1}\fi \endgroup}
\def\mn@eprint#1#2{\mn@eprint@#1:#2::\@nil}
\def\mn@eprint@arXiv#1{\href {http://arxiv.org/abs/#1} {{\tt arXiv:#1}}}
\def\mn@eprint@dblp#1{\href {http://dblp.uni-trier.de/rec/bibtex/#1.xml} {dblp:#1}}
\def\mn@eprint@#1:#2:#3:#4\@nil{\def\@tempa {#1}\def\@tempb {#2}\def\@tempc {#3}\ifx \@tempc \@empty \let \@tempc \@tempb \let \@tempb \@tempa \fi \ifx \@tempb \@empty \def\@tempb {arXiv}\fi \@ifundefined {mn@eprint@\@tempb}{\@tempb:\@tempc}{\expandafter \expandafter \csname mn@eprint@\@tempb\endcsname \expandafter{\@tempc}}}

\bibitem[\protect\citeauthoryear{Abdurro'uf et~al.,}{Abdurro'uf et~al.}{2022}]{Abdurrouf+22}
Abdurro'uf et~al., 2022, The {{Seventeenth Data Release}} of the {{Sloan Digital Sky Surveys}}: {{Complete Release}} of {{MaNGA}}, {{MaStar}} and {{APOGEE-2 Data}} (\mn@eprint {arxiv} {2112.02026}), \mn@doi{10.3847/1538-4365/ac4414}

\bibitem[\protect\citeauthoryear{Aguado et~al.,}{Aguado et~al.}{2019}]{Aguado+19}
Aguado D.~S.,  et~al., 2019, \mn@doi [The Astrophysical Journal Supplement Series] {10.3847/1538-4365/aaf651}, 240, 23

\bibitem[\protect\citeauthoryear{{Astropy Collaboration} et~al.,}{{Astropy Collaboration} et~al.}{2013}]{astropy:2013}
{Astropy Collaboration} et~al., 2013, \mn@doi [\aap] {10.1051/0004-6361/201322068}, \href {https://ui.adsabs.harvard.edu/abs/2013A&A...558A..33A} {558, A33}

\bibitem[\protect\citeauthoryear{{Baker}, {Davis}, {van de Voort}  \& {Ruffa}}{{Baker} et~al.}{2024}]{max_baker}
{Baker} M.~K.,  {Davis} T.~A.,  {van de Voort} F.,   {Ruffa} I.,  2024, \mn@doi [arXiv e-prints] {10.48550/arXiv.2412.03707}, \href {https://ui.adsabs.harvard.edu/abs/2024arXiv241203707B} {p. arXiv:2412.03707}

\bibitem[\protect\citeauthoryear{{Baldwin}, {Phillips}  \& {Terlevich}}{{Baldwin} et~al.}{1981}]{BPT}
{Baldwin} J.~A.,  {Phillips} M.~M.,   {Terlevich} R.,  1981, \mn@doi [\pasp] {10.1086/130766}, \href {https://ui.adsabs.harvard.edu/abs/1981PASP...93....5B} {93, 5}

\bibitem[\protect\citeauthoryear{{Bassani}, {Dadina}, {Maiolino}, {Salvati}, {Risaliti}, {Della Ceca}, {Matt}  \& {Zamorani}}{{Bassani} et~al.}{1999}]{bassani}
{Bassani} L.,  {Dadina} M.,  {Maiolino} R.,  {Salvati} M.,  {Risaliti} G.,  {Della Ceca} R.,  {Matt} G.,   {Zamorani} G.,  1999, \mn@doi [\apjs] {10.1086/313202}, \href {https://ui.adsabs.harvard.edu/abs/1999ApJS..121..473B} {121, 473}

\bibitem[\protect\citeauthoryear{{Bassett}, {Bekki}, {Cortese}  \& {Couch}}{{Bassett} et~al.}{2017}]{bassett17}
{Bassett} R.,  {Bekki} K.,  {Cortese} L.,   {Couch} W.,  2017, \mn@doi [\mnras] {10.1093/mnras/stx958}, \href {https://ui.adsabs.harvard.edu/abs/2017MNRAS.471.1892B} {471, 1892}

\bibitem[\protect\citeauthoryear{Belfiore et~al.,}{Belfiore et~al.}{2019}]{Belfiore+19}
Belfiore F.,  et~al., 2019, \mn@doi [The Astronomical Journal] {10.3847/1538-3881/ab3e4e}, 158, 160

\bibitem[\protect\citeauthoryear{{Bertola}, {Buson}  \& {Zeilinger}}{{Bertola} et~al.}{1992}]{bertola}
{Bertola} F.,  {Buson} L.~M.,   {Zeilinger} W.~W.,  1992, \mn@doi [\apjl] {10.1086/186675}, \href {https://ui.adsabs.harvard.edu/abs/1992ApJ...401L..79B} {401, L79}

\bibitem[\protect\citeauthoryear{Blanton et~al.,}{Blanton et~al.}{2017}]{Blanton+17}
Blanton M.~R.,  et~al., 2017, \mn@doi [The Astronomical Journal] {10.3847/1538-3881/aa7567}, 154, 28

\bibitem[\protect\citeauthoryear{Bryant et~al.,}{Bryant et~al.}{2015}]{SAMI_swoi}
Bryant J.~J.,  et~al., 2015, \mn@doi [Monthly Notices of the Royal Astronomical Society] {10.1093/mnras/stu2635}, 447, 2857

\bibitem[\protect\citeauthoryear{Bryant et~al.,}{Bryant et~al.}{2018}]{bryant_croom}
Bryant J.~J.,  et~al., 2018, \mn@doi [Monthly Notices of the Royal Astronomical Society] {10.1093/mnras/sty3122}, 483, 458

\bibitem[\protect\citeauthoryear{Bundy et~al.,}{Bundy et~al.}{2015}]{Bundy+15a}
Bundy K.,  et~al., 2015, \mn@doi [Astrophysical Journal] {10.1088/0004-637X/798/1/7}, \href {http://adsabs.harvard.edu/abs/2015ApJ...798....7B} {798, 7}

\bibitem[\protect\citeauthoryear{Capelo \& Dotti}{Capelo \& Dotti}{2016}]{capelo}
Capelo P.~R.,  Dotti M.,  2016, \mn@doi [Monthly Notices of the Royal Astronomical Society] {10.1093/mnras/stw2872}, 465, 2643

\bibitem[\protect\citeauthoryear{Cherinka et~al.,}{Cherinka et~al.}{2019}]{Cherinka+19}
Cherinka B.,  et~al., 2019, \mn@doi [The Astronomical Journal] {10.3847/1538-3881/ab2634}, 158, 74

\bibitem[\protect\citeauthoryear{{Choi}, {Somerville}, {Ostriker}, {Hirschmann}  \& {Naab}}{{Choi} et~al.}{2024}]{choi}
{Choi} E.,  {Somerville} R.~S.,  {Ostriker} J.~P.,  {Hirschmann} M.,   {Naab} T.,  2024, \mn@doi [\apj] {10.3847/1538-4357/ad245a}, \href {https://ui.adsabs.harvard.edu/abs/2024ApJ...964...54C} {964, 54}

\bibitem[\protect\citeauthoryear{Combes}{Combes}{2001}]{combes}
Combes F.,  2001, in , Advanced Lectures on the Starburst-AGN Connection.
World Scientific, pp 223--278

\bibitem[\protect\citeauthoryear{Croom et~al.,}{Croom et~al.}{2021}]{croom}
Croom S.~M.,  et~al., 2021, \mn@doi [Monthly Notices of the Royal Astronomical Society] {10.1093/mnras/stab229}, 505, 991–1016

\bibitem[\protect\citeauthoryear{Davies et~al.,}{Davies et~al.}{2014}]{davies_maciejewski}
Davies R.~I.,  et~al., 2014, \mn@doi [The Astrophysical Journal] {10.1088/0004-637x/792/2/101}, 792, 101

\bibitem[\protect\citeauthoryear{{Davis} \& {Bureau}}{{Davis} \& {Bureau}}{2016}]{davis_bureau}
{Davis} T.~A.,  {Bureau} M.,  2016, \mn@doi [\mnras] {10.1093/mnras/stv2998}, \href {https://ui.adsabs.harvard.edu/abs/2016MNRAS.457..272D} {457, 272}

\bibitem[\protect\citeauthoryear{Davis et~al.,}{Davis et~al.}{2011a}]{davis_alatalo}
Davis T.~A.,  et~al., 2011a, \mn@doi [Monthly Notices of the Royal Astronomical Society] {10.1111/j.1365-2966.2011.19355.x}, 417, 882

\bibitem[\protect\citeauthoryear{Davis et~al.,}{Davis et~al.}{2011b}]{tim_davis}
Davis T.~A.,  et~al., 2011b, \mn@doi [Monthly Notices of the Royal Astronomical Society] {10.1111/j.1365-2966.2011.19355.x}, 417, 882

\bibitem[\protect\citeauthoryear{Domínguez Sánchez, Margalef, Bernardi  \& Huertas-Company}{Domínguez Sánchez et~al.}{2021}]{sanchez}
Domínguez Sánchez H.,  Margalef B.,  Bernardi M.,   Huertas-Company M.,  2021, \mn@doi [Monthly Notices of the Royal Astronomical Society] {10.1093/mnras/stab3089}, 509, 4024–4036

\bibitem[\protect\citeauthoryear{Drory et~al.,}{Drory et~al.}{2015}]{Drory+15}
Drory N.,  et~al., 2015, \mn@doi [The Astronomical Journal] {10.1088/0004-6256/149/2/77}, 149, 77

\bibitem[\protect\citeauthoryear{Fan et~al.,}{Fan et~al.}{2016}]{fan}
Fan L.,  et~al., 2016, \mn@doi [The Astrophysical Journal Letters] {10.3847/2041-8205/822/2/l32}, 822, L32

\bibitem[\protect\citeauthoryear{{Gallagher}, {Faber}  \& {Balick}}{{Gallagher} et~al.}{1975}]{gallagher}
{Gallagher} J.~S.,  {Faber} S.~M.,   {Balick} B.,  1975, \mn@doi [\apj] {10.1086/153948}, \href {https://ui.adsabs.harvard.edu/abs/1975ApJ...202....7G} {202, 7}

\bibitem[\protect\citeauthoryear{Gobat et~al.,}{Gobat et~al.}{2018}]{Gobat_2018}
Gobat R.,  et~al., 2018, \mn@doi [Nature Astronomy] {10.1038/s41550-017-0352-5}, 2, 239–246

\bibitem[\protect\citeauthoryear{Gunn et~al.,}{Gunn et~al.}{2006}]{Gunn+06}
Gunn J.~E.,  et~al., 2006, \mn@doi [The Astronomical Journal] {10.1086/500975}, 131, 2332

\bibitem[\protect\citeauthoryear{{Harrison} \& {Ramos Almeida}}{{Harrison} \& {Ramos Almeida}}{2024}]{2024Galax..12...17H}
{Harrison} C.~M.,  {Ramos Almeida} C.,  2024, \mn@doi [Galaxies] {10.3390/galaxies12020017}, \href {https://ui.adsabs.harvard.edu/abs/2024Galax..12...17H} {12, 17}

\bibitem[\protect\citeauthoryear{{Hopkins}, {Hernquist}, {Cox}, {Younger}  \& {Besla}}{{Hopkins} et~al.}{2008}]{Hopkins2008}
{Hopkins} P.~F.,  {Hernquist} L.,  {Cox} T.~J.,  {Younger} J.~D.,   {Besla} G.,  2008, \mn@doi [\apj] {10.1086/592087}, \href {https://ui.adsabs.harvard.edu/abs/2008ApJ...688..757H} {688, 757}

\bibitem[\protect\citeauthoryear{{Hoyle} \& {Fowler}}{{Hoyle} \& {Fowler}}{1963}]{hoyle}
{Hoyle} F.,  {Fowler} W.~A.,  1963, \mn@doi [\nat] {10.1038/197533a0}, \href {https://ui.adsabs.harvard.edu/abs/1963Natur.197..533H} {197, 533}

\bibitem[\protect\citeauthoryear{{Ilha} et~al.,}{{Ilha} et~al.}{2019}]{ilha19}
{Ilha} G.~S.,  et~al., 2019, \mn@doi [\mnras] {10.1093/mnras/sty3373}, \href {https://ui.adsabs.harvard.edu/abs/2019MNRAS.484..252I} {484, 252}

\bibitem[\protect\citeauthoryear{{Ilha}, {Krabbe}, {Riffel}, {Dors}, {Riffel}, {Rembold}, {Storchi-Bergmann}  \& {Mallmann}}{{Ilha} et~al.}{2024}]{ilha}
{Ilha} G.~S.,  {Krabbe} A.~C.,  {Riffel} R.~A.,  {Dors} O.~L.,  {Riffel} R.,  {Rembold} S.~B.,  {Storchi-Bergmann} T.,   {Mallmann} N.~D.,  2024, \mn@doi [\mnras] {10.1093/mnras/stae1685}, \href {https://ui.adsabs.harvard.edu/abs/2024MNRAS.532.2988I} {532, 2988}

\bibitem[\protect\citeauthoryear{Jin et~al.,}{Jin et~al.}{2016}]{jin}
Jin Y.,  et~al., 2016, \mn@doi [Monthly Notices of the Royal Astronomical Society] {10.1093/mnras/stw2055}, 463, 913

\bibitem[\protect\citeauthoryear{{Kannappan} \& {Fabricant}}{{Kannappan} \& {Fabricant}}{2001}]{kannappan01}
{Kannappan} S.~J.,  {Fabricant} D.~G.,  2001, \mn@doi [\aj] {10.1086/318027}, \href {https://ui.adsabs.harvard.edu/abs/2001AJ....121..140K} {121, 140}

\bibitem[\protect\citeauthoryear{{Kocevski}, {Brightman}, {Koekemoer}  \& {Nandra}}{{Kocevski} et~al.}{2015}]{Kocevski2015}
{Kocevski} D.~D.,  {Brightman} M.,  {Koekemoer} A.~M.,   {Nandra} K.,  2015, {Are Compton-Thick AGN the Missing Link Between Mergers and Black Hole Growth?}, HST Proposal. Cycle 22, ID. \#13868

\bibitem[\protect\citeauthoryear{{Kormendy} \& {Ho}}{{Kormendy} \& {Ho}}{2013}]{kormendy}
{Kormendy} J.,  {Ho} L.~C.,  2013, \mn@doi [\araa] {10.1146/annurev-astro-082708-101811}, \href {https://ui.adsabs.harvard.edu/abs/2013ARA&A..51..511K} {51, 511}

\bibitem[\protect\citeauthoryear{{Kormendy} \& {Richstone}}{{Kormendy} \& {Richstone}}{1995}]{kormendy_richstone}
{Kormendy} J.,  {Richstone} D.,  1995, \mn@doi [\araa] {10.1146/annurev.aa.33.090195.003053}, \href {https://ui.adsabs.harvard.edu/abs/1995ARA&A..33..581K} {33, 581}

\bibitem[\protect\citeauthoryear{Lagos et~al.,}{Lagos et~al.}{2017}]{lagos}
Lagos C. d.~P.,  et~al., 2017, \mn@doi [Monthly Notices of the Royal Astronomical Society] {10.1093/mnras/stx2667}, 473, 4956

\bibitem[\protect\citeauthoryear{Lamastra, Bianchi, Matt, Perola, Barcons  \& Carrera}{Lamastra et~al.}{2009}]{lamastra}
Lamastra A.,  Bianchi S.,  Matt G.,  Perola G.~C.,  Barcons X.,   Carrera F.~J.,  2009, \mn@doi [Astronomy &amp; Astrophysics] {10.1051/0004-6361/200912023}, 504, 73–79

\bibitem[\protect\citeauthoryear{Law et~al.,}{Law et~al.}{2015}]{Law+15}
Law D.~R.,  et~al., 2015, \mn@doi [The Astronomical Journal] {10.1088/0004-6256/150/1/19}, 150, 19

\bibitem[\protect\citeauthoryear{Law et~al.,}{Law et~al.}{2016}]{Law+16}
Law D.~R.,  et~al., 2016, \mn@doi [The Astronomical Journal] {10.3847/0004-6256/152/4/83}, 152, 83

\bibitem[\protect\citeauthoryear{Law et~al.,}{Law et~al.}{2021}]{Law+21}
Law D.~R.,  et~al., 2021, \mn@doi [The Astronomical Journal] {10.3847/1538-3881/abcaa2}, 161, 52

\bibitem[\protect\citeauthoryear{{Lees}, {Knapp}, {Rupen}  \& {Phillips}}{{Lees} et~al.}{1991}]{Lees}
{Lees} J.~F.,  {Knapp} G.~R.,  {Rupen} M.~P.,   {Phillips} T.~G.,  1991, \mn@doi [\apj] {10.1086/170494}, \href {https://ui.adsabs.harvard.edu/abs/1991ApJ...379..177L} {379, 177}

\bibitem[\protect\citeauthoryear{Li et~al.,}{Li et~al.}{2023}]{zhi_li}
Li Z.,  et~al., 2023, How Nested Bars Enhance, Modulate, and are Destroyed by Gas Inflows (\mn@eprint {arXiv} {2310.04666})

\bibitem[\protect\citeauthoryear{Lianou, Xilouris, Madden  \& Barmby}{Lianou et~al.}{2016}]{lianou}
Lianou S.,  Xilouris E.,  Madden S.~C.,   Barmby P.,  2016, \mn@doi [Monthly Notices of the Royal Astronomical Society] {10.1093/mnras/stw1467}, 461, 2856

\bibitem[\protect\citeauthoryear{{Lynden-Bell} \& {Rees}}{{Lynden-Bell} \& {Rees}}{1971}]{lynden-bell}
{Lynden-Bell} D.,  {Rees} M.~J.,  1971, \mn@doi [\mnras] {10.1093/mnras/152.4.461}, \href {https://ui.adsabs.harvard.edu/abs/1971MNRAS.152..461L} {152, 461}

\bibitem[\protect\citeauthoryear{{Macchetto}, {Marconi}, {Axon}, {Capetti}, {Sparks}  \& {Crane}}{{Macchetto} et~al.}{1997}]{macchetto}
{Macchetto} F.,  {Marconi} A.,  {Axon} D.~J.,  {Capetti} A.,  {Sparks} W.,   {Crane} P.,  1997, \mn@doi [\apj] {10.1086/304823}, \href {https://ui.adsabs.harvard.edu/abs/1997ApJ...489..579M} {489, 579}

\bibitem[\protect\citeauthoryear{{Mart{\'\i}n-Navarro}, {Shankar}  \& {Mezcua}}{{Mart{\'\i}n-Navarro} et~al.}{2022}]{navarro}
{Mart{\'\i}n-Navarro} I.,  {Shankar} F.,   {Mezcua} M.,  2022, \mn@doi [\mnras] {10.1093/mnrasl/slab112}, \href {https://ui.adsabs.harvard.edu/abs/2022MNRAS.513L..10M} {513, L10}

\bibitem[\protect\citeauthoryear{McConnell \& Ma}{McConnell \& Ma}{2013}]{mcconnell}
McConnell N.~J.,  Ma C.-P.,  2013, \mn@doi [The Astrophysical Journal] {10.1088/0004-637x/764/2/184}, 764, 184

\bibitem[\protect\citeauthoryear{Negri, Ciotti  \& Pellegrini}{Negri et~al.}{2014}]{negri}
Negri A.,  Ciotti L.,   Pellegrini S.,  2014, \mn@doi [Monthly Notices of the Royal Astronomical Society] {10.1093/mnras/stt2505}, 439, 823

\bibitem[\protect\citeauthoryear{{Peterson}}{{Peterson}}{1997}]{peterson}
{Peterson} B.~M.,  1997, {An Introduction to Active Galactic Nuclei}

\bibitem[\protect\citeauthoryear{Pizzella, Corsini, Vega~Beltrán  \& Bertola}{Pizzella et~al.}{2004}]{pizzella}
Pizzella A.,  Corsini E.~M.,  Vega~Beltrán J.~C.,   Bertola F.,  2004, \mn@doi [Astronomy &amp; Astrophysics] {10.1051/0004-6361:20047183}, 424, 447–454

\bibitem[\protect\citeauthoryear{{Raimundo}}{{Raimundo}}{2021}]{raimundo21}
{Raimundo} S.~I.,  2021, \mn@doi [\aap] {10.1051/0004-6361/202040248}, \href {https://ui.adsabs.harvard.edu/abs/2021A&A...650A..34R} {650, A34}

\bibitem[\protect\citeauthoryear{{Raimundo}, {Davies}, {Canning}, {Celotti}, {Fabian}  \& {Gandhi}}{{Raimundo} et~al.}{2017}]{raimundo17}
{Raimundo} S.~I.,  {Davies} R.~I.,  {Canning} R.~E.~A.,  {Celotti} A.,  {Fabian} A.~C.,   {Gandhi} P.,  2017, \mn@doi [\mnras] {10.1093/mnras/stw2635}, \href {https://ui.adsabs.harvard.edu/abs/2017MNRAS.464.4227R} {464, 4227}

\bibitem[\protect\citeauthoryear{Raimundo, Malkan  \& Vestergaard}{Raimundo et~al.}{2023}]{sandra}
Raimundo S.~I.,  Malkan M.,   Vestergaard M.,  2023, \mn@doi [Nature Astronomy] {10.1038/s41550-022-01880-z}, 7, 463–472

\bibitem[\protect\citeauthoryear{{Reines} \& {Volonteri}}{{Reines} \& {Volonteri}}{2015}]{reines}
{Reines} A.~E.,  {Volonteri} M.,  2015, \mn@doi [\apj] {10.1088/0004-637X/813/2/82}, \href {https://ui.adsabs.harvard.edu/abs/2015ApJ...813...82R} {813, 82}

\bibitem[\protect\citeauthoryear{{Rembold} et~al.,}{{Rembold} et~al.}{2017}]{rembold}
{Rembold} S.~B.,  et~al., 2017, \mn@doi [\mnras] {10.1093/mnras/stx2264}, \href {https://ui.adsabs.harvard.edu/abs/2017MNRAS.472.4382R} {472, 4382}

\bibitem[\protect\citeauthoryear{{Rembold} et~al.,}{{Rembold} et~al.}{2024}]{Rembold+24}
{Rembold} S.~B.,  et~al., 2024, \mn@doi [\mnras] {10.1093/mnras/stad3584}, \href {https://ui.adsabs.harvard.edu/abs/2024MNRAS.527.6722R} {527, 6722}

\bibitem[\protect\citeauthoryear{Riess et~al.,}{Riess et~al.}{2022}]{riess}
Riess A.~G.,  et~al., 2022, \mn@doi [The Astrophysical Journal Letters] {10.3847/2041-8213/ac5c5b}, 934, L7

\bibitem[\protect\citeauthoryear{Riffel et~al.,}{Riffel et~al.}{2021}]{Riffel+21}
Riffel R.,  et~al., 2021, \mn@doi [Monthly Notices of the Royal Astronomical Society] {10.1093/mnras/staa3907}, 501, 4064

\bibitem[\protect\citeauthoryear{{Riffel} et~al.,}{{Riffel} et~al.}{2023}]{riffel}
{Riffel} R.,  et~al., 2023, \mn@doi [\mnras] {10.1093/mnras/stad2234}, \href {https://ui.adsabs.harvard.edu/abs/2023MNRAS.524.5640R} {524, 5640}

\bibitem[\protect\citeauthoryear{{Riffel} et~al.,}{{Riffel} et~al.}{2024}]{Riffel+24}
{Riffel} R.,  et~al., 2024, \mn@doi [\mnras] {10.1093/mnras/stae1192}, \href {https://ui.adsabs.harvard.edu/abs/2024MNRAS.531..554R} {531, 554}

\bibitem[\protect\citeauthoryear{Ristea et~al.,}{Ristea et~al.}{2022}]{ristea}
Ristea A.,  et~al., 2022, \mn@doi [Monthly Notices of the Royal Astronomical Society] {10.1093/mnras/stac2839}, 517, 2677

\bibitem[\protect\citeauthoryear{{Rosario} et~al.,}{{Rosario} et~al.}{2018}]{Rosario}
{Rosario} D.~J.,  et~al., 2018, \mn@doi [\mnras] {10.1093/mnras/stx2670}, \href {https://ui.adsabs.harvard.edu/abs/2018MNRAS.473.5658R} {473, 5658}

\bibitem[\protect\citeauthoryear{Schawinski, Koss, Berney  \& Sartori}{Schawinski et~al.}{2015}]{schawinski2015activegalacticnucleiflicker}
Schawinski K.,  Koss M.,  Berney S.,   Sartori L.,  2015, Active galactic nuclei flicker: an observational estimate of the duration of black hole growth phases of ~1e5 years (\mn@eprint {arXiv} {1505.06733}), \url {https://arxiv.org/abs/1505.06733}

\bibitem[\protect\citeauthoryear{Shlosman, Begelman  \& Frank}{Shlosman et~al.}{1990}]{shlosman}
Shlosman I.,  Begelman M.,   Frank J.,  1990, Nature, 345, 679

\bibitem[\protect\citeauthoryear{Smee et~al.,}{Smee et~al.}{2013}]{Smee+13}
Smee S.~A.,  et~al., 2013, \mn@doi [The Astronomical Journal] {10.1088/0004-6256/146/2/32}, 146, 32

\bibitem[\protect\citeauthoryear{{Storchi-Bergmann} \& {Schnorr-M{\"u}ller}}{{Storchi-Bergmann} \& {Schnorr-M{\"u}ller}}{2019}]{2019NatAs...3...48S}
{Storchi-Bergmann} T.,  {Schnorr-M{\"u}ller} A.,  2019, \mn@doi [Nature Astronomy] {10.1038/s41550-018-0611-0}, \href {https://ui.adsabs.harvard.edu/abs/2019NatAs...3...48S} {3, 48}

\bibitem[\protect\citeauthoryear{Taylor, Federrath  \& Kobayashi}{Taylor et~al.}{2018}]{taylor}
Taylor P.,  Federrath C.,   Kobayashi C.,  2018, \mn@doi [Monthly Notices of the Royal Astronomical Society] {10.1093/mnras/sty1439}, 479, 141–152

\bibitem[\protect\citeauthoryear{{Thakar} \& {Ryden}}{{Thakar} \& {Ryden}}{1996}]{thakar}
{Thakar} A.~R.,  {Ryden} B.~S.,  1996, \mn@doi [\apj] {10.1086/177037}, \href {https://ui.adsabs.harvard.edu/abs/1996ApJ...461...55T} {461, 55}

\bibitem[\protect\citeauthoryear{Urrutia, Becker, White, Glikman, Lacy, Hodge  \& Gregg}{Urrutia et~al.}{2008}]{urrutia}
Urrutia T.,  Becker R.~H.,  White R.~L.,  Glikman E.,  Lacy M.,  Hodge J.~A.,   Gregg M.~D.,  2008, The Astrophysical Journal, 698, 1095

\bibitem[\protect\citeauthoryear{Wake et~al.,}{Wake et~al.}{2017}]{Wake+17}
Wake D.~A.,  et~al., 2017, \mn@doi [The Astronomical Journal] {10.3847/1538-3881/aa7ecc}, 154, 86

\bibitem[\protect\citeauthoryear{Westfall et~al.,}{Westfall et~al.}{2019}]{Westfall+19}
Westfall K.~B.,  et~al., 2019, \mn@doi [The Astronomical Journal] {10.3847/1538-3881/ab44a2}, 158, 231

\bibitem[\protect\citeauthoryear{{Woltjer}}{{Woltjer}}{1959}]{woltjer}
{Woltjer} L.,  1959, \mn@doi [\apj] {10.1086/146694}, \href {https://ui.adsabs.harvard.edu/abs/1959ApJ...130...38W} {130, 38}

\bibitem[\protect\citeauthoryear{Wu \& Shen}{Wu \& Shen}{2022}]{wu}
Wu Q.,  Shen Y.,  2022, \mn@doi [The Astrophysical Journal Supplement Series] {10.3847/1538-4365/ac9ead}, 263, 42

\bibitem[\protect\citeauthoryear{Yan et~al.,}{Yan et~al.}{2016a}]{Yan+16a}
Yan R.,  et~al., 2016a, \mn@doi [The Astronomical Journal] {10.3847/0004-6256/151/1/8}, 151, 8

\bibitem[\protect\citeauthoryear{Yan et~al.,}{Yan et~al.}{2016b}]{Yan+16}
Yan R.,  et~al., 2016b, \mn@doi [The Astronomical Journal] {10.3847/0004-6256/152/6/197}, 152, 197

\bibitem[\protect\citeauthoryear{Young et~al.,}{Young et~al.}{2011}]{Young}
Young L.~M.,  et~al., 2011, \mn@doi [Monthly Notices of the Royal Astronomical Society] {10.1111/j.1365-2966.2011.18561.x}, 414, 940

\bibitem[\protect\citeauthoryear{{van de Voort}, {Davis}, {Kere{\v{s}}}, {Quataert}, {Faucher-Gigu{\`e}re}  \& {Hopkins}}{{van de Voort} et~al.}{2015}]{vandevoort15}
{van de Voort} F.,  {Davis} T.~A.,  {Kere{\v{s}}} D.,  {Quataert} E.,  {Faucher-Gigu{\`e}re} C.-A.,   {Hopkins} P.~F.,  2015, \mn@doi [\mnras] {10.1093/mnras/stv1217}, \href {https://ui.adsabs.harvard.edu/abs/2015MNRAS.451.3269V} {451, 3269}

\makeatother
\end{thebibliography}





\bsp	
\label{lastpage}
\end{document}